\documentstyle[12pt]{article}
\begin{document}
\vspace*{2cm}
\begin{center}
{\LARGE\bf Effective Action for Multi-Regge  }
\end{center}
\begin{center}
{\LARGE\bf Processes in QCD }
\end{center}
\vspace{0.5cm}
\begin{center}
{\large\bf R. Kirschner}\\{\bf DESY - Institut f\"ur
Hochenergiephysik, Zeuthen, Germany}
\end{center}
\begin{center}
{\large\bf L.N. Lipatov\footnote{
Supported by the Russian Fond of Fundamental Research,
Grant No 93-02-16809}}\\
{\bf St. Petersburg Nucl. Phys. Institute}\\
{\bf Gatchina, Russia} \\
{\bf and} \\
{\bf Universit\"at-GH-Siegen, Siegen, Germany}
\end{center}
\begin{center}
{\large\bf L. Szymanowski\footnote{ Supported in part by the
Polish Goverment
Grant and by the German-Polish
Scientific and \newline \hspace*{0.6cm}Technological Agreement. }}\\
{\bf Soltan Inst. for Nuclear
Studies}\\{\bf Warsaw, Poland}
\end{center}
\vspace*{2.0cm}
We construct the effective Lagrangian describing QCD in the
 multi-Regge kinematics. It is obtained from the original QCD
 Lagrangian by eliminating modes of gluon and quark fields not
 appearing in this underlying kinematics.

\newpage
\section{Introduction}

Two types of evolution equations for partonic distributions in
the small $x$ deep-inelastic ep-scattering are used now. The
GLAP equation \cite{1} is applied for finding the
$Q^2$-dependence of structure functions and the BFKL equation \cite{2}
is applied for fixing their $x$ dependence. These equations
correspond to summing the leading logarithmic terms $\sim
(g^2\ln Q^2)^n$ and $\sim (g^2\ln \frac{1}{x})^n$
correspondingly. Their solutions grow rapidly in the limit $x
\to 0$ and violate the Froissart bound for total cross-sections
\cite{2}. The violation is a consequence of the fact that the
scattering amplitudes in the leading logarithmic approximation
(LLA) do not satisfy $s$-channel unitarity constraints.
Equations for the unitarity corrections have been investigated
in \cite{3}. The treatment of the multi-Reggeon exchanges turns
out to be difficult besides of special cases \cite{4} or of the
limit of large gauge group ($N_c \to \infty$) \cite{5}. A new
method for unitarizing the results of LLA was suggested by one
of the authors \cite{6}. It is based on an effective field
theory coinciding with QCD in the case, when all particles in
the intermediate states of $s$- and $u$- channels have the
multi-Regge kinematics \cite{2}(see Fig.~1)
\begin{eqnarray}
  s &=& (\mbox{p}_A + \mbox{p}_B)^2 \;,\;\;\;
  s_i = (\mbox{k}_i + \mbox{k}_{i-1})^2 \;,
          \qquad i = 1, \ldots , n+1 \nonumber \\
  \mbox{k}_0 &\equiv& \mbox{p}_{A'}\;, \qquad \mbox{k}_{n+1} \equiv
          \mbox{p}_{B'}\;, \qquad \mbox{k}_i =
          \mbox{q}_{i} - \mbox{q}_{i+1} \;,
                  \qquad t_i = \mbox{q}^2_i \;,
          \nonumber  \\
        \mbox{k}^\mu &=&
        \frac{(\mbox{p}_A \mbox{k})}{(\mbox{p}_A \mbox{p}_B)}
        \mbox{p}_B^\mu +
        \frac{(\mbox{p}_B \mbox{k})}{(\mbox{p}_A \mbox{p}_B)}
        \mbox{p}_A^\mu + k_{\bot}^\mu \;,\qquad k^2_{\perp} =
-{\vec k}^2_{\perp} \nonumber \\
  s &\gg& s_1, \sim s_2, \sim \ldots ,\sim s_{n+1} \gg |t_1| \sim
       |t_2| \sim \ldots \sim |t_{n+1}| \nonumber \\
  s_1 s_2 &\ldots& s_{n+1} = s \prod^n_{i=1} {\vec k}^2_{\perp i}
   \nonumber \\
  &\mbox{k}_1\mbox{p}_A& \ll \mbox{k}_2\mbox{p}_A \ll \ldots \ll
  \mbox{k}_n\mbox{p}_A     \nonumber \\
  &\mbox{k}_1\mbox{p}_B& \gg \mbox{k}_2\mbox{p}_B \gg \ldots \gg
  \mbox{k}_n\mbox{p}_B
\end{eqnarray}
The corresponding effective action was constructed in
ref.\cite{6} only for pure gluodynamics and for gravity.
Recently another effective theory for the high energy scattering
in the Yang-Mills model was suggested by H. and E. Verlinde
\cite{7}. The effective action for the gravity, suggested in
\cite{6}, was used by D. Amati, M. Ciafaloni and G. Veneziano
for calculation of the scattering amplitudes at super-Plankian
energies \cite{8}.

The multi-Regge effective action can be viewed as the action
reproducing the multiparticle tree amplitudes in the kinematics
(1) in the most simple way. This can be taken as a guiding
principle to obtain the action as discussed in \cite{6} and in
\cite{9}.

In this paper we present the direct way from the original QCD
action to the effective action. We start from the QCD action in
the light-cone gauge. The gluon and quark fields are separated
according to regions in the momentum space determined by the
kinematics (1). Modes with highly virtual momenta are excluded
by means of equations of motion. We distinguish the modes with
momenta corresponding to the kinematics of the quanta scattered
(and produced) or exchanged in the peripheral scattering process.
A discussion of the effective action emphasizing symmetry
properties will be given in a further publication \cite{10}.

\section{The QCD action in the light-cone gauge}

For simplicity we consider massless quarks and suppress the
flavour indices for quark fields $\psi(x)$. In this case the QCD
Lagrangian takes the form
\begin{eqnarray}
{\cal L} &=& - \frac{1}{4}G^a_{\mu\nu}G^{a\mu\nu} +
i{\bar\psi}\gamma^{\mu}(\partial_{\mu} - igt^aA^a_{\mu})\psi
\nonumber \\
G^a_{\mu\nu} &=& \partial_{\mu}A^a_{\nu} - \partial_{\nu}A^a_{\mu}
+ gf^{abc}A^b_{\mu}A^c_{\nu}
\end{eqnarray}
where $f^{abc}$ are the structure constants, $t^a$ is the
generator of the gauge group in quark representation ($[t^a,t^b]
= if^{abc}t^c$) and $g$ is the QCD coupling constant.

To derive the effective action for multi-Regge processes it is
convenient to fix the gauge for the vector potential
$A^a_{\mu}$. In the center-of-mass system, where
\begin{equation}
\mbox{p}^0_A = \mbox{p}^0_B = \mbox{p}^3_A = -\mbox{p}^3_B =
\frac{\sqrt{s}}{2} ,\qquad p_{A\perp} = p_{B\perp} = 0
\end{equation}
for initial mass-less particles with momenta p$_A$ and p$_B$,
 we chose the light-cone gauge (cf. \cite{6})
\begin{equation}
   A_- \equiv A_0 - A_3 = \mbox{p}^{\mu}_B A_{\mu} = 0
\end{equation}

In this gauge  action (2) depends on $A_+ \equiv A_0 + A_3$
only quadratically and therefore one can exclude $A_+$ from (2)
using equations of motion
\begin{equation}
  A^a_+ =
-\frac{1}{\partial_-}\left(\partial_{\rho}A^{a\rho}\right) -
g\frac{1}{\partial_-^2}\left[ iA_{\rho}T^a(\partial_-A^{\rho}) -
\frac{1}{2}{\bar \psi}_-\gamma_-t^a\psi_- \right]
\end{equation}
where $\rho = 1,2\;,\qquad \partial_- = \frac{\partial}{\partial
x^-} \;,\qquad x^- = x^0 - x^3$ and $T^a$ denotes the generator of
gauge group in the adjoint representation ($(T^a)_{bc} =
-if^{abc}$) .

The field $\psi$ can be written as the sum of two components $\psi_{\pm}$
\begin{equation}
   \psi = \psi_+ + \psi_- , \qquad \psi_{\pm} = P_{\pm} \psi , \qquad
   P_{\pm} = \frac{1}{4}\gamma_{\mp}\gamma_{\pm} \; ,
\end{equation}
where $P_{\pm}$ are projectors into corresponding subspaces. In
the gauge (4) equations of motion for $\psi_+ , {\bar \psi}_+$
are easily solved
\begin{eqnarray}
   \psi_+ &=& - \frac{1}{4}\frac{1}{\partial_-}\gamma_-
   \left({\hat \partial} - ig{\hat {\hat A}}\right) \psi_-
   \nonumber \\
    {\bar \psi}_+ &=& - \frac{1}{4}\frac{1}{\partial_-}\left(
   (\partial^{\rho}{\bar\psi}_-)\gamma_{\rho} +
   ig{\bar\psi}_-\hat{\hat{A}}\right)\gamma_- \nonumber \\
   \hat{\hat{A}} &\equiv& A^a_{\sigma}\gamma^{\sigma}t^a ,
    \qquad {\hat {\partial}} \equiv
    \gamma_{\sigma}\partial^{\sigma}
\end{eqnarray}
Putting $A_+$, $\psi_+$, ${\bar\psi}_+$ from (5), (7) in
Lagrangian (2) we obtain
\begin{eqnarray}
{\cal L}^{QCD} &=& \frac{1}{2} A^a_\sigma \Box A^{a \sigma} -
ig (\partial_-
  A_\sigma) T^a A_\sigma (\frac{1}{\partial_-}
  \partial_\varrho A^{a\varrho}) \nonumber \\
  && - \frac{g^2}{2} (\partial_- A_\sigma) T^a A^\sigma
  \frac{1}{\partial^2_-}
  (\partial_- A_\rho) T^a A^\rho - ig (\partial_\varrho A^a_\sigma)
  A^\varrho T^a A^\sigma \nonumber \\
  && + \frac{i}{4} \bar{\psi}_- \frac{\gamma_-}{\partial_-} \Box \psi_- -
    \frac{g}{2} \bar{\psi}_- \gamma_- t^a \psi_- (\frac{1}{\partial_-}
        \partial_\varrho A^{a\varrho}) + \nonumber \\
  && + \frac{g^2}{8} \bar{\psi}_- \gamma_- t^a \psi_-
  \frac{1}{\partial^2_-}
    \bar{\psi}_- \gamma_- g^a \psi_- + \frac{ig^2}{2} \bar{\psi}_- \gamma_-
        t^a \psi_- \frac{1}{\partial^2_-}
        (\partial_- A_\varrho) T^a A^\varrho
        \nonumber \\
  && - \frac{g}{4} \bar{\psi}_- \hat{\partial}
  \frac{1}{\partial_-} \gamma_-
    \hat{\hat{A}} \psi_- - \frac{g}{4} \bar{\psi}_-
    \hat{\hat{A}}\frac{1}{\partial_-}\gamma_- \hat{\partial}
        \psi_- + \nonumber \\
  && + \frac{ig^2}{4} \bar{\psi}_- \hat{\hat{A}} \frac{1}{\partial_-}
  \gamma_-\hat{\hat{A}}\psi_-
     + \frac{g^2}{4} A_\varrho T^a A_\sigma A^\varrho T^a A^\sigma
\end{eqnarray}
where $\Box = 4\partial_+\partial_- +
\partial_{\sigma}\partial^{\sigma}$ is the d'Alembertian.

For high-energy processes the last term in the Lagrangian (8) is
not essential (Ref.\cite{2}). It does not give also any
contribution to the effective action. This fact is in accordance
with Ref.\cite{7}. However it turnes out that the contributions
from the transverse part $\sim
\partial_{\sigma}\partial^{\sigma}$ of the d'Alembertian $\Box$
in the first term and from the three-linear fourth term of
Lagrangian (8) are needed to reproduce correctly the effective
action of Ref.\cite{6}. This seems to be in conflict with the
arguments of Ref.\cite{7} about the possibility to neglect at
high energies the term $-
\frac{1}{4}G^a_{\rho\sigma}G^{a\rho\sigma}$ with all transverse indices.

\section{Exclusion of heavy modes in the multi-Regge kinematics}

It is known, that to obtain the generating functional for scattering
amplitudes in the tree approximation one has to calculate the Lagrangian
on the
solutions $\tilde{A}_\sigma$ of equations of motion having a prescribed
asymptotic behaviour at $t \rightarrow \pm \infty$ \cite{11}.
In this section
we use the equations of motions in order to exclude from the
action (4) the fields
describing strongly virtual gluons and quarks (heavy modes)
appearing in the
Feynman diagrams for production amplitudes in the multi-Regge
kinematics (1).
One diagram of this type is shown in Fig.~2 (where $k^2
\simeq k^2_\| =
s_2 \gg |k^2_{\perp}|)$. We solve the equations of motion for
heavy modes within the
perturbation theory. The following analysis is based on the
first perturbative
order. At the end of this chapter we comment on possible
improvements of this
approximation.

Let us decompose each of $A_\sigma, \psi_-$ and $\bar{\psi}_-$ in
the sums
of two fields for the strongly $(s)$ and moderately $(m)$ virtual
particles:
\begin{equation}
  A_\sigma = A^{(s)}_\sigma + A^{(m)}_\sigma\;\;,\;\; \psi_- =
  \psi^{(s)}_- +
    \psi^{(m)}\;\;,\;\; \bar{\psi}_- = \bar{\psi}^{(s)}_- +
    \bar{\psi}^{(m)}_- .
\end{equation}
In the kinetic terms of Lagrangian (8) one can neglect the
interference
contributions between $s$- and $m$-fields with a good accuracy:
  \begin{equation}
   {\cal L}^{ kin} \cong  2 A^{(s)}_\sigma \partial_+
     \partial_- A^{(s)\sigma} + \frac{1}{2} A^{(m)}_\sigma \Box
         A^{(m) \sigma} +
         i \bar{\psi}^{(s)}_- \gamma_- \partial_+
         \psi^{(s)}_- \\
  + \frac{i}{4} \bar{\psi}^{(m)}_- \frac{\gamma_-}{\partial_-}
     \Box \psi^{(m)}_-
  \end{equation}
  where we took into account, that the $s$-field virtuality
  is large only due
  to the presence of the large longitudinal momenta of external
  particles
  $(\Box \simeq 4 \partial_+ \partial_-)$(see Fig. 2).

  The interaction terms for $s$ fields are obtained from Lagrangian (8)
  by keeping (after decomposition (9)) the most important (in the underlying
  kinematics) terms. These terms are enhanced by
  the operator $\frac{1}{\partial_-}$ acting on the $m$ field with the
  smallest $k_-$ momentum component. Up to the first order in $g$
  the essential interaction terms involving $s$ fields are the
  following (cf. (10))
  \begin{eqnarray}
    {\cal L}^{(s)} &=& 2 A^{(s)a}_\varrho \partial_+ \partial_- A^{(s) a
          \varrho} + i \bar{\psi}^{(s)}_- \gamma_- \partial_+ \psi^{(s)}_-
          \nonumber \\
     && +g [i A_\sigma T^a \partial_- A^\sigma - \frac{1}{2} \bar{\psi}_-
          \gamma_- t^a \psi_-] (\frac{1}{\partial_-}
          \partial_\varrho A^{(m)a
          \varrho}) \\
        && + \frac{g}{4} (\partial^\sigma \bar{\psi}^{(m)}_-) \gamma_\sigma
          \frac{1}{\partial_-} \gamma_- \hat{\hat{A}} \psi_- -\frac{g}{4}
          \bar{\psi}_- \hat{\hat{A}} \frac{1}{\partial_-}
          \gamma_- \hat{\partial}
          \psi^{(m)}_- \nonumber
\end{eqnarray}
where the fields $A_\sigma, \psi_-$ and $\bar{\psi}_-$ are decomposed
according to (9). Let us note that because of the absence of the
above mentioned
enhancement factor a contribution to (11) of fourth term in Lagrangian (8)
as well as all the remaining terms can be omitted in our
approximation (they
involve only $m$ fields, see eq.~(13)).

The $s$ field solutions of equations of motions following from
Lagrangian (11)
can be easily found in the perturbation theory
\begin{eqnarray}
  \tilde{A}^{(s)a}_\sigma &=& - \frac{ig}{2} A^{(m)}_\sigma T^a
     (\frac{1}{\partial_+ \partial_-} \partial_\sigma A^{(m) \sigma}) -
          \nonumber \\
  && -\frac{g}{16} \left[ \left( \frac{1}{\partial_+ \partial_-}
     \bar{\psi}^{(m)}_- \right) \hat{\partial} \gamma_- t^a \gamma_\sigma
         \frac{1}{\partial_-} \psi^{(m)}_- - \left( \frac{1}{\partial_-}
         \bar{\psi}^{(m)}_- \right) t^a \gamma_\sigma \gamma_-
         \frac{\hat{\partial}}{\partial_- \partial_+} \psi^{(m)}_- \right]
         \nonumber \\
  && + {\cal O}(g^2) \nonumber \\
  \tilde{\psi}^{(s)}_- &=& - \frac{ig}{2} \left\{ t^a \psi^{(m)}_-
      (\frac{1}{\partial_+ \partial_-} \partial_\varrho A^{(m) a \varrho})
          + \frac{1}{2} \hat{\hat{A}}^{(m)}
          (\frac{\hat{\partial}}{\partial_+
          \partial_-} \psi^{(m)}_-) \right \} + {\cal O}(g^2)\\
 \bar{\tilde{\psi}}^{(s)}_- &=& \frac{ig}{2} \left\{ (\frac{1}{\partial_+
     \partial_-} \partial_\varrho A^{(m)a\varrho}) \bar{\psi}^{(m)}_- t^a
         - \frac{1}{2} (\frac{\partial^\varrho}{\partial_+ \partial_-}
         \bar{\psi}^{(m)}_-) \gamma_\varrho \hat{\hat{A}}^{(m)} \right\}
         + {\cal O}(g^2)\nonumber
\end{eqnarray}
where we took into account that in the multi-Regge kinematics (1), for each
pair of fields, the field having a smaller value of $\partial_-$, should have
a bigger value of $\partial_+$.

We substitute the fields $A^{s)}, \psi^{(s)}, \bar{\psi}^{(s)}$ in eq.~(11)
by their classical counterparts (12), and replace the part ${\cal L}^{(s)}$
(11) of the Yang-Mills action (8) by this result.
We obtain a modified action involving only  the
$m$-fields:
 \begin{equation}
   {\cal L}^{mod} = {\cal L}^{QCD}
   |_{A,\psi,\bar{\psi}\rightarrow A^{(m)},
   \psi^{(m)}, \bar{\psi}^{(m)}} + \Delta {\cal L} \; ,
 \end{equation}
 where (cf. (11))
 \begin{equation}
   \Delta {\cal L} = -2 \tilde{A}^{(s)}_\sigma \partial_+ \partial_-
   \tilde{A}^{(s)}_\sigma -
   i \bar{\tilde{\psi}}^{(s)}_- \gamma_- \partial_+
   \tilde{\psi}^{(s)} + {\cal O}(g^3) \; .
 \end{equation}
 Performing the differentiations after substitution of eqs.~(12) in eq.~(14)
 we have to take into account which field carry small $k_-$ (large $k_+$)
 momentum components. Thus, the differentiations $\partial_+$
 and $\partial_-$
 in (14) give an essential contribution only when they are applied to the
 fields with the factor $\frac{1}{\partial_+}$ and without this factor,
 correspondingly. Taking into account also the freedom of integrating
 by parts,
the  expression (14) can be represented as follows (cf.~(11)).
 \begin{eqnarray}
 \Delta {\cal L}
 &=& \frac{g}{4} [A^{(m)}_\varrho T^a (\partial_- A^{(m)\varrho}) +
     \frac{i}{2} \bar{\psi}^{(m)}_- t^a \gamma_- \psi^{(m)}_- ]
         (\frac{1}{\partial_+\partial_-}
         \partial_\sigma A^{(m)\sigma} ) T^a
         (\frac{1}{\partial_-} \partial_\eta A^{(m)\eta}) + \nonumber \\
 &+& \frac{ig^2}{8} (\frac{1}{\partial_+ \partial_-}
    \partial_\sigma A^{(m)\sigma}) [(\frac{\partial^\varrho}{\partial_-}
        \bar{\psi}^{(m)}_-) \gamma_\varrho \gamma_- t^a \hat{\hat{A}}
        \psi^{(m)}_- - \bar{\psi}^{(m)} \hat{\hat{A}} t^a \gamma_-
        \hat{\partial}
        \frac{1}{\partial_-} \psi^{(m)}_-]
\end{eqnarray}
All other terms, including those from four quark fields are negligible in
the multi-Regge kinematics.
In particular, we neglect the contribution of the term
symmetric to the one
written in eq.~(15) in which both fermions carry small $k_-$
momentum components
since the matrix element involving fermions is small.

In the next section we shall simplify the action (13) for
the $m$ fields by
the use of their equations of motion.
 Above we excluded the heavy fields
from the action approximately
up to the order $ g^2$ of the perturbation theory (see (15)), but it
can be done in all orders because ${\cal L}^{(s)}$ in eq.(11)
depends only
quadratically on them. Instead of eq.~(12) we would obtain expressions,
containing in particular the path ordered exponentials $ P \exp
(\frac{i}{2} \int_L A^a_+ T^a dx^+)$ with $A_+$ given by eq.~(5) and
therefore ${\cal L}^{mod}$ in eq.(13) generally is nonlocal and contains
strongly nonlinear interaction terms. Nevertheless, the effective action,
which is constructed in the next sections, remains to be three-linear in
fields even after taking into account the higher order corrections to the
classical solutions (12) for heavy fields. We hope to discuss this point
 elsewhere.

\section{Equations of motion for Coulomb fields}

After eliminating the heavy modes we separate the
remaining modes $A^{(m)},
\psi^{(m)}$ of the gluon and quark fields into a part
involving modes $A',
\psi'$ with the momenta obeying $|k_+ k_-| \ll |k^2_\perp|$ and a part
$A, \psi$ with the momenta obeying $|k_+ k_-| \simeq |k^2_\perp|$. We call
the modes $A', \psi'$ Coulombic; in their kinetic terms the longitudinal
derivatives can be omitted: $\Box \rightarrow \partial_\sigma
\partial^\sigma$ and  therefore
 these modes describe the instantaneous Coulomb interactions.

The additional terms (15) in the Lagrangian obtained by elimination of the
heavy modes can be factorized into triple vertices connected by Coulombic
propagators. In this way we see that the terms (15) are
reproduced if we add
to the original action (8) the following  induced terms
\begin{equation}
  \tilde{\cal L}^{mod} = {\cal L}^{QCD} + \Delta {\cal L}^{ind}
\end{equation}
where
\begin{eqnarray}
  \Delta {\cal L}^{ind} &=& \frac{ig}{4} (\partial_- \partial_\varrho
     A'^{a\varrho}) \left( \frac{1}{\partial_+ \partial_-}
     \partial_\sigma A^\sigma \right) T^a \left( \frac{1}{\partial_-}
          \partial_\eta A^\eta \right) + \nonumber \\
  && + \frac{g}{8} \left( \frac{1}{\partial_+ \partial_-}
     \partial_\sigma  A^{a\sigma} \right) \left[ (\partial^\varrho
         \bar{\psi}'_-) \gamma_\varrho \gamma_- t^a
     (\frac{\hat{\partial}}{\partial_-} \psi_-) + \right. \\
  && \left. + \left( \frac{\partial^\varrho}{\partial_-} \bar{\psi}_-
     \right) \gamma_\varrho \gamma_- t^a \hat{\partial} \psi'_- \right]
         \nonumber
\end{eqnarray}
Let us verify, that the quartic terms (15) can be obtained from
(16) by the use of equations of motions for Coulomb fields:
\begin{eqnarray}
  \partial_\sigma \partial^\sigma \tilde{A}'^{a\varrho} &=& ig
     \frac{\partial^\varrho}{\partial_-}((\partial_- A_\sigma)
     T^a A^\sigma) + \frac{g}{2} \frac{\partial^\varrho}{\partial_-}
         (\bar{\psi}_- \gamma_- t^a \psi_-) \nonumber \\
   && - \frac{ig}{4} \partial_- \partial^\varrho ( (
      \frac{1}{\partial_+ \partial_-} \partial_\sigma A^\sigma) T^a
          (\frac{1}{\partial_-} \partial_\eta A^\eta)) \nonumber \\
   \frac{\partial_\sigma \partial^\sigma}{\partial_-} \tilde{\psi}'_-
      &=& ig \frac{\hat{\partial}}{\partial_-} (\hat{\hat{A}}
      \psi_-) + \frac{ig}{2} \hat{\partial} \left( t^a
          (\frac{\hat{\partial}}{\partial_-}
      \psi_-)(\frac{1}{\partial_+ \partial_-} \partial_\varrho
      A^{a \varrho})\right) \\
   \frac{\partial_\sigma \partial^\sigma}{\partial_-}
   \bar{\tilde{\psi}}'_-
      &=& - ig (\frac{\partial^\sigma}{\partial_-} \bar{\psi}_-
      \hat{\hat{A}}) \gamma_\sigma - \frac{ig}{2} \partial^\sigma
          \left( (\frac{1}{\partial_+ \partial_-}
          \partial_\varrho A^{a\varrho})
          (\frac{\partial^\eta}{\partial_-} \bar{\psi}_-)
          \right) \gamma_\eta
          t^a \gamma_\sigma \nonumber
\end{eqnarray}

In  (18) we write down only the terms, which arise from
varying the fields containing factors $\frac{1}{\partial_-}$ in
${\cal L}^{QCD}$ (8) and nonsingular ones for $\partial_-
\rightarrow 0$ in $\Delta {\cal L}^{ind}$ (16). These fields
describe correspondingly the emission and absorption of Coulomb
particles in the crossing channel.

Putting solutions $\tilde{A}, \tilde{\psi}_-,
\bar{\tilde{\psi}}_-$ of eqs.~(18) back in
$\tilde{\cal{L}}^{mod}$ we obtain several four-linear terms:
\begin{eqnarray}
&&  - \frac{1}{2} \tilde{A}'^a_\sigma \partial_\varrho \partial^\varrho
   \tilde{A}'^{a\sigma} - \frac{i}{4} \bar{\tilde{\psi}}'_- \gamma_-
   \frac{\partial_\varrho \partial^\varrho}{\partial_-}
   \tilde{\psi}'_- = \nonumber \\
 &=& - \frac{g^2}{2} [ i (\partial_- A_\sigma) T^a
 A^\sigma + \frac{1}{2}
   \bar{\psi}_- \gamma_- t^a \psi_-] \frac{1}{\partial^2_-} [i (\partial_-
   A_\varrho) T^a A^\varrho + \frac{1}{2} \bar{\psi}_-
   \gamma_- t^a \psi_-] \nonumber \\
 && - \frac{ig^2}{4} \bar{\psi}_- \hat{\hat{A}}
 \frac{\gamma_-}{\partial_-}
   \hat{\hat{A}} \psi_- + \frac{ig^2}{8}
   (\frac{1}{\partial_+ \partial_-}
   \partial_\sigma A^{a\sigma}) (\frac{\partial^\varrho}{\partial_-}
   \bar{\psi}_-) \gamma_\varrho t^a \gamma_- \hat{\hat{A}}
   \psi_- \nonumber \\
 && - \frac{ig^2}{8} (\frac{1}{\partial_+ \partial_-} \partial_\sigma
    A^{a\sigma}) \bar{\psi}_- \hat{\hat{A}} \gamma_- t^a
    (\frac{\hat{\partial}}{\partial_-} \psi_-) + \nonumber \\
 && + \frac{ig^2}{4} [i (\partial_- A_\sigma) T^a A^\sigma + \frac{1}{2}
    \bar{\psi}_- \gamma_- t^a \psi_-] (\frac{1}{\partial_+ \partial_-}
        \partial_\varrho A^\varrho) T^a (\frac{1}{\partial_-}
        \partial_\eta
        A^\eta) \nonumber \\
 && + \ldots
 \end{eqnarray}
 The last three terms in eq. (19) reproduce the contribution
(15) of heavy modes and first two are cancelled with the
four-linear singular terms in eq.~(8). Neglecting the last
nonsingular term in (8) one can write instead of (13) the
following comparatively simple expression for the effective lagrangian:
\begin{eqnarray}
  {\cal L}^{eff} &=& \frac{1}{2} A^a_\varrho \Box A^{a\varrho} +
       \frac{i}{4} \bar{\psi}_- \frac{\gamma_-}{\partial_-} \Box
       \psi_-   \nonumber \\
    && - g A'^a_+ [ i A_\sigma T^a (\partial_- A^\sigma) - \frac{1}{2}
           \bar{\psi}_- \gamma_- t^a \psi_-] \nonumber \\
        && + g [ \bar{\psi}'_+ \hat{\hat{A}} \psi_-
        + \bar{\psi}_- \hat{\hat{A}}
           \psi'_+] - ig  (\partial_\varrho A^a_\sigma) A^\varrho
       T^a A^\sigma - \nonumber \\
        && - \frac{ig}{8} (\partial_\sigma \partial^\sigma A'^a_-)
           (\frac{1}{\partial_+ \partial_-} \partial_\varrho A^\varrho)
           T^a (\frac{1}{\partial_-} \partial_\eta A^\eta) - \\
        && - \frac{g}{2} (\frac{1}{\partial_+ \partial_-}
        \partial_\sigma
           A^{a\sigma}) [(\partial^\varrho \bar{\psi}'_-)
       \gamma_\varrho t^a \psi_+ + \bar{\psi}_+ t^a \hat{\partial}
           \psi'_-] \nonumber
\end{eqnarray}
where we introduced the following notations (cf.~(5),(7)) for
the Coulomb fields (in the corresponding kinematics)
\begin{equation}
  \begin{array}{ll}
    A'_+ = - \frac{1}{\partial_-} \partial_\sigma A'^\sigma
               & A'_- = -2 \frac{\partial_-
               \partial_\sigma}{\partial_\varrho
                     \partial^\varrho} A'^\sigma \\
    \psi'_+ = - \frac{1}{4} \frac{1}{\partial_-} \gamma_- \hat{\partial}
               \psi'_- & \bar{\psi}'_+ = - \frac{1}{4}
           (\frac{\partial^\sigma}{\partial_-}
           \bar{\psi}'_-) \gamma_\sigma
                   \gamma_-
        \end{array}
\end{equation}
Let us note that the role of $A'_-$ field  in the fermionic
case is played by $\psi'_-$. We want also to emphasize that
fields appearing in (20) without prime contain all modes of moderately
virtual fields, i.e. they are in fact $A + A'$ or $\psi_- + \psi'_-$.

Although $A'_+, \psi'_+, \bar{\psi}'_+$ according
to eqs. (21) are expressed
through $A'_-, \psi'_-, \bar{\psi}'_-$, we shall consider them in the
following discussion as independent fields,
supposing in particular, that
\begin{equation}
  0 = \langle A'_+ (x) A'_+(0) \rangle = \langle A'_- (x) A'_- (0)
  \rangle
    = \langle {\psi}'_-(x)      \bar{\psi}'_-(0)\rangle = \langle
    {\psi}'_+(x) \bar{\psi}'_+(0)\rangle
\end{equation}
for the virtual (Coulomb) fields. The vanishing of the
 Green functions for
the same fields (22) is needed to put effectively to zero the
two superfluous
terms in the expression (19) in the second order of perturbation
theory, which
are cancelled with the corresponding nonlinear terms in eq.~(8).

In accordance with eqs. (21) we should take the following free
actions for the Coulomb fields:
\begin{equation}
  {\cal L}^{free}_{Coul} = \frac{1}{2} A'^a_+ \partial_\varrho
    \partial^\varrho A'^a_- + i \bar{\psi}'_- \hat{\partial} \psi'_+
        + i \bar{\psi}'_+ \hat{\partial} \psi'_-
\end{equation}
Note that to reproduce correctly the Coulomb field propagators the
coefficients in (23) are chosen to be different
 from what one would obtain by formal
substitution of (21). In writing eq. (23) we took also into
account that field $A'_+$ contains only the longitudinal part
$\sim \partial_{\sigma}$ of
field $A'_\sigma$ (eq.~(21)).

In the next sections the action (20) will be transformed to a
simpler form by introducing separately two sorts of fields for
the produced and virtual particles.

\section{Interacting vertices for virtual and real particles}

The three-linear terms in Lagrangian (20) describe the various
processes of particle production and scattering. The real
(produced) and virtual (Coulomb) particles have completely
different kinematics and therefore it is natural to introduce
for them different notations \cite{6}. Temporarily we leave for the
fields, corresponding to produced particles the old notations
$A_\sigma, \psi_-, \bar{\psi}_-$ and use the notations (21) for
fields, describing Coulomb-like forces.

All  vertices for the `forward' scattering off right
particles  linear in $A'_+, \psi'_+$ and $\bar{\psi}'_+$
 can be written down from (20):
\begin{eqnarray}
  {\cal L}^{(R)}_{scat} &=& -g [i A_\sigma T^a (\partial_- A^\sigma)
    - \frac{1}{2} \bar{\psi}_- \gamma_- t^a \psi_-] A'^a_+
     \nonumber \\
  && + g (\bar{\psi}'_+ \hat{\hat{A}} \psi_- + \bar{\psi}_-
      \hat{\hat{A}} \psi'_+)
\end{eqnarray}
In the dominant kinematics these vertices describe the scattering
of particles close in momentum to p$_A$ off the right
particles close
 in momentum to p$_B$.

Since $A'^a_\pm$ contains $\partial_\varrho A'^{a \varrho}$ and do
not contain $\varepsilon_{\varrho\sigma} \partial^\varrho A'^{a
\sigma}$ one can use one of the following substitutions for the
fields describing the Coulomb particles:
\begin{equation}
  A'^{a \varrho} \rightarrow \frac{1}{\partial_\eta \partial^\eta}
  \partial^\varrho \partial_\sigma A'^{a \sigma} = - \frac{1}{2}
  \frac{\partial^\varrho}{\partial_-} A'^a_- \;\; , \;
  - \frac{1}{\partial_\eta
  \partial^\eta} \partial_- \partial^\varrho A'^a_+
\end{equation}
The production vertices ${\cal L}_{prod}$ are obtained with the
use of (25) from the terms giving contributions
to ${\cal L}^{(R)}_{scat}$
(24) and from the three last terms in eq.~(20)
\begin{eqnarray}
  {\cal L}_{prod}
  &=& g A'^a_+ [ i A_\sigma T^a (\partial^\sigma A'_-) + \frac{1}{2}
     \bar{\psi}'_- \gamma_- t^a \psi_- + \frac{1}{2}
     \bar{\psi}_- \gamma_- t^a \psi'_-] \nonumber \\
  && + g [\bar{\psi}'_+ \hat{\hat{A}} \psi'_- +  \bar{\psi}'_-
     \hat{\hat{A}} \psi'_+ - \frac{1}{2} \bar{\psi}'_+ t^a
         (\hat{\partial} \frac{1}{\partial_-} A'^a_-) \psi_- -
         \frac{1}{2} \bar{\psi}_- t^a (\hat{\partial}
         \frac{1}{\partial_-}
         A'^a_-) \psi'_+] \nonumber \\
  && + \frac{ig}{4} (\partial_\varrho \partial^\varrho A'^a_-)
     (\frac{1}{\partial_+ \partial_-} \partial_\sigma A^\sigma) T^a
         A'_+ \nonumber \\
  && - \frac{g}{2} [ \bar{\psi}'_+ t^a (\frac{1}{\partial_+ \partial_-}
    \partial_\sigma A^{a\sigma}) \hat{\partial} \psi'_- - \bar{\psi}'_-
        \hat{\partial} t^a (\frac{1}{\partial_+ \partial_-}
        \partial_\sigma
        A^{a\sigma}) \psi'_+ ]\nonumber \\
  && - \frac{g}{2} [(\frac{1}{\partial_+} \bar{\psi}_+) t^a A'^a_+
    \hat{\partial} \psi'_- - \bar{\psi}'_- \hat{\partial} t^a A'^a_+
        (\frac{1}{\partial_+} \psi_+)]
\end{eqnarray}
We shall transform ${\cal L}_{prod}$ to a  simpler form in
the next section. Now, let us consider the forward scattering
off left particles. Here one should take into account that
according to our definitions (21) $A'_-, \psi'_-$ and
$\bar{\psi}'_-$ do not contain the singularities
$\frac{1}{\partial_-}$ and therefore for the forward scattering off
left particles all three-linear terms of ${\cal L}^{eff}$ (20)
give contributions of the necessary order:
\begin{eqnarray}
  {\cal L}^{(L)}_{scat}
  &=& \frac{g}{2} (\frac{1}{\partial_-} \partial_\varrho A^{a \varrho})
    [i (\partial_\sigma A'_-) T^a A^\sigma + i (\partial_- A_\sigma)
        T^a (\frac{\partial^\sigma}{\partial_-} A'_-) - \bar{\psi}'_-
        \gamma_- t^a \psi_- \nonumber \\
        && \mathop{\phantom{free horizontal space,
        free horizontal space}}
          - \bar{\psi}_- \gamma_- t^a \psi'_- ] \nonumber \\
        && +g [ \bar{\psi}_+ \hat{\hat{A}} \psi'_- + \bar{\psi}'_-
          \hat{\hat{A}} \psi_+ - \frac{1}{2} \bar{\psi}_+
      (\hat{\partial} \frac{1}{\partial_-}
          A'^a_-) t^a \psi_- - \frac{1}{2} \bar{\psi}_-
      (\hat{\partial} \frac{1}{\partial_-} A'^a_-) t^a \psi_+]
          \nonumber \\
   && - \frac{ig}{8} (\partial_\eta \partial^\eta A'^a_-)
      (\frac{1}{\partial_+ \partial_-} \partial_\sigma A^\sigma) T^a
          (\frac{1}{\partial_-} \partial_\varrho A^\varrho)
      \nonumber \\
   && - \frac{g}{2} [(\partial^\varrho \bar{\psi}'_-) \gamma_\varrho
      t^a (\frac{1}{\partial_+ \partial_-} \partial_\sigma A^{a\sigma})
          \psi_+ + \bar{\psi}_+ t^a (\frac{1}{\partial_+ \partial_-}
          \partial_\sigma A^{a\sigma}) \hat{\partial} \psi'_-]
      \nonumber \\
   && + \frac{ig}{2} \left( (\partial_\varrho A^a_\sigma) -
      (\partial_\sigma A^a_\varrho) \right)
      (\frac{\partial^\varrho}{\partial_-} A'_-) T^a A^\sigma
\end{eqnarray}
We stress that the last term in expression (27) appears from
the nonsingular Yang-Mills vertex which disappears in
Verlinde scaling limit \cite{7}. One can transform the expression (27) by
integrating by part to the form:
\newpage
\begin{eqnarray}
 {\cal L}^{(L)}_{scat}
   &=& ig (\partial_\sigma A'^a_-) A^\sigma T^a (\frac{1}{\partial_-}
       \partial_\varrho A^\varrho) + \frac{ig}{4} (\frac{1}{\partial_-}
           A'^a_-) [(\partial_\varrho \partial^\varrho A_\sigma) T^a
       A^\sigma - A^\sigma T^a (\partial_\varrho
      \partial^\varrho A_\sigma)] \nonumber \\
  && - \frac{ig}{8} (\partial_\sigma \partial^\sigma A'^a_-)
       (\frac{1}{\partial_+ \partial_-} \partial_\varrho
       A^\varrho) T^a (\frac{1}{\partial_-} \partial_\eta A^\eta) +
           \frac{g}{2} A'^a_- \bar{\psi}_+ \gamma_+ t^a \psi_+
           \nonumber \\
   &&  +\frac{g}{2} (\frac{\partial^\varrho}{\partial_+\partial_-}
       \partial_\sigma A^{a\sigma})[\bar{\psi}_+ \gamma_\varrho t^a
           \psi'_- + \bar{\psi}'_- t^a \gamma_\varrho \psi_+]
       \nonumber \\
   &&  + g [ \bar{\psi}_+ \hat{\hat{A}} \psi'_- + \bar{\psi}'_-
       \hat{\hat{A}} \psi_+]
\end{eqnarray}
Note, that there are also the three-linear vertices with only
Coulomb fields. In particular, from three last (induced) terms
in eq.~(20) one obtains the following contributions for two
right $(+)$ and one left $(-)$ fields:
\begin{eqnarray}
  {\cal L}_{-++}
  &=& - \frac{ig}{8} (\partial_\varrho \partial^\varrho A'^a_-)
       (\frac{1}{\partial_+} A'_+)T^a A'_+ + \nonumber \\
  && + \frac{g}{2} [(\partial^\varrho \bar{\psi}'_-)
      \gamma_\varrho t^a \psi'_+ + \bar{\psi}'_+ t^a
      (\hat{\partial} \psi'_-)] (\frac{1}{\partial_+} A'^a_+)
\end{eqnarray}
The symmetric expression for the interaction of two left $(-)$ and
one right $(+)$ fields is
\begin{eqnarray}
  {\cal L}_{+--}
  &=& -  \frac{ig}{8} (\partial_\varrho \partial^\varrho A'^a_+)
       (\frac{1}{\partial_-} A'_-) T^a A'_- + \nonumber \\
  && + \frac{g}{2} [(\partial^\varrho \bar{\psi}'_+) \gamma_\varrho
      t^a \psi'_- + \bar{\psi}'_- t^a (\hat{\partial} \psi'_+)]
          (\frac{1}{\partial_-} A'^a_- )
\end{eqnarray}
With the help of integration by part and keeping in mind that the
$k_-$ component
of $A'_+$ is smaller than those of the fields $A'_-$ one can rewrite
eq.~(30) in the form
\begin{eqnarray}
  {\cal L}_{+--} &=& - \frac{ig}{4} A'^a_+
       (\frac{\partial^\varrho}{\partial_-}  A'_-) T^a (\partial_\varrho
           A'_-) \nonumber \\
  && - \frac{g}{2} [\bar{\psi}'_+ t^a (\frac{\hat{\partial}}{\partial_-}
       A'^a_-) \psi'_- +  \bar{\psi}'_- t^a
       (\frac{\hat{\partial}}{\partial_-}  A'^a_-) \psi'_+ ]
       \nonumber \\
  &&  - \frac{ig}{4} A'^a_+ (\frac{1}{\partial_-} A'_-) T^a
        (\partial^\varrho \partial_\varrho A'_-) \nonumber \\
  &&  - \frac{g}{2} [ \bar{\psi}'_+ t^a (\hat{\partial} \psi'_-)
      + (\partial_\varrho \bar{\psi}'_-) \gamma^\varrho t^a \psi'_+]
          (\frac{1}{\partial_-} A'^a_-)
\end{eqnarray}
The first half of expression (31) is obtained from the terms in eq.~(20)
containing fields $A'_+, \psi'_+, \bar{\psi}'_+$. The second
half of it should be obtained from induced terms in eq.~(20)
after a more accurate calculation of them (earlier it was
assumed that derivatives $\partial_+$ from the fields $A'_-, \psi'_-,
\bar{\psi}'_-$ are negligible which is not the case now).

\section{Effective action for multi-Regge processes in  QCD}
It is convenient to introduce the complex and light-cone
coordinates in the two-dimensional transverse and longitudinal
subspaces, correspondingly
  \begin{equation}
     \begin{array}{llll}
           \varrho = x^1 + ix^2, & \varrho^* = x^1 - ix^2, &
       \partial = \frac{\partial}{\partial\varrho} , & \partial^*
            = \frac{\partial}{\partial \varrho^*}  \\
        x^\pm = x^0 \pm x^3  &  \partial_\pm =
       \frac{\partial}{\partial z^\pm}  &  &
           \end{array}
\end{equation}
and the analogous components for transverse fields
and $\gamma$-matrices:
\begin{equation}
  \begin{array}{ll}
    A = A^1 + iA^2, &  A^* = A^1 - i A^2  \\
        \gamma = \gamma^1 + i\gamma^2, & \gamma^* =
        \gamma^1 - i \gamma^2 \\
        \gamma_\pm = \gamma_0 \pm \gamma_3 &
 \end{array}
 \end{equation}
 In accordance with ref. \cite{6} we describe the produced particles by
 the complex scalar fields related with $A$ by the equation:
  \begin{equation}
    A = i \partial^* \phi, A^* = - i \partial \phi^*
\end{equation}
For the Coulomb fields we use instead of $A'_\pm$ the old denotions
$A_\pm$:
 \begin{equation}
   A'_\pm \rightarrow A_\pm
 \end{equation}
 Let us introduce the following basis for the spinors:
\begin{equation}
  u_{--} = \frac{1}{2} \left( \begin{array}{r}
      0 \\ 1 \\ 0 \\ -1  \end{array} \right) ,
  u_{-+} = \frac{(-1)}{2} \left( \begin{array}{l}
      1 \\ 0 \\ 1 \\ 0   \end{array}  \right) ,
  u_{+-} = \frac{1}{2} \left(  \begin{array}{l}
      0 \\ 1 \\ 0 \\ 1 \end{array} \right) ,
  u_{++} = \frac{1}{2} \left( \begin{array}{r}
     1 \\ 0 \\ -1 \\ 0  \end{array} \right) ,
\end{equation}
satisfying the relations:
\begin{eqnarray}
   && u_{-i} = \frac{1}{4} \gamma_+ \gamma_- u_{-i},
      \quad \gamma_+ u_{-i} = 0,  \nonumber \\
  &&  u_{+i} = \frac{1}{4} \gamma_- \gamma_+ u_{+i}, \quad
      \gamma_- u_{+i} = 0 , \nonumber \\
  &&  u_{i+} = \frac{1}{4} \gamma \gamma^* u_{i+}, \quad
      \gamma u_{i+} = 0 , \\
  &&  u_{i-} = - \frac{1}{4} \gamma^* \gamma u_{i-} , \quad
      \gamma^* u_{i-}  = 0  , \nonumber \\
  &&  u_{++} = - \frac{1}{2} \gamma_- u_{-+} = \frac{1}{2}
      \gamma u_{+-} ,\nonumber \\
  &&  u_{--} = \frac{1}{2} \gamma_+ u_{+-} = -\frac{1}{2}
      \gamma^* u_{-+} \nonumber
\end{eqnarray}
and the normalization conditions:
\begin{equation}
  1 = \bar{u}_{+i} \gamma_+ u_{+i} = \bar{u}_{-i} \gamma_- u_{-i} =
    2 \bar{u}_{+-} u_{--} = -2 \bar{u}_{++} u_{-+} = -2 \bar{u}_{-+}
        u_{++} = 2\bar{u}_{--} u_{+-}
\end{equation}
Then the fermions are described by anticommuting fields
$\chi_\pm,  \chi^*_\pm$ and $a_\pm, a^*_\pm$ related with initial
fields $\psi_\pm, \psi'_\pm$ written in the above basis as
follows:
\begin{eqnarray}
  \psi_\pm &=& 2i [(\partial \chi_\pm) u_{\pm +} - (\partial^*
      \chi^*_\pm) u_{\pm -} ] \nonumber \\
  \psi'_\pm &=& a^*_\pm u_{\pm +}  + a_\pm u_{\pm -}
\end{eqnarray}
We remind, that the fields $a_+, a^*_+$ and $a_-, a^*_-$
are considered as independent ones. For the complex
conjugated fields one obtains from eqs.~(39):
\begin{eqnarray}
 \bar{\psi}_\pm &=& -2i [(\partial^* \bar{\chi}_\pm)
     \bar{u}_{\pm +} - (\partial \bar{\chi}^*_\pm) \bar{u}_{\pm
     -}], \nonumber \\
  \bar{\psi}'_\pm &=& \bar{a^*}_\pm \bar{u}_{\pm +} +
      \bar{a}\pm \bar{u}_{\pm -},
\end{eqnarray}
where by the definition the complex conjugation is denoted
by bar. Only for the Majorana particles one obtains
$\bar{\chi} = \chi^* \; , \; \bar{a} = a^*$.

The newly introduceD fields $\chi_+, \chi^*_+$ are expressed
through the fields $\chi_-, \chi^*_-$ (see (7)).
\begin{equation}
  \chi_+ = - \frac{\partial^*}{\partial_-} \chi^*_- , \quad \chi^*_+ =
       - \frac{\partial}{\partial_-} \chi_-
\end{equation}
{}From eqs. (20) and (23) using above definitions we obtain
the free Lagrangian for the effective theory in the form
\begin{eqnarray}
{\cal L}_{free} & = & - 2A^a_+\partial\partial^* A^a_- - i[
\bar{a}^{a*}_+(\partial a^a_-) + \bar{a}^a_+(\partial^* a^{a*}_-) +
\bar{a}^{a*}_-(\partial a^a_+) + \nonumber \\
& + & \bar{a}^a_-(\partial^* a^{a*}_+)] -
\frac{1}{2}(\partial^*\phi^a)
\Box (\partial\phi^{a *}) + \nonumber \\
& + & i [ (\partial^*\bar{\chi}^a_-)\frac{\Box}{\partial_-}
(\partial \chi^a_-) + (\partial\bar{\chi}^{a *}_-)
\frac{\Box}{\partial_-}
(\partial^*\chi^{a*}_-)]
\end{eqnarray}
{}From eq.(24) one can find the following Lagrangian for
scattering off right particles:
\begin{eqnarray}
   {\cal L}^{(R)}_{scat} &=& - \frac{ig}{2}A^a_+        \left[
(\partial_- \partial^*\phi) T^a(\partial \phi^*) + (\partial_-
\partial \phi^*) T^a(\partial^* \phi)\right] \nonumber \\
  &+& 2gA^a_+ \left[(\partial^*\bar{\chi}_-)t^a(\partial\chi_-)
+ (\partial\bar{\chi}^*_-)t^a(\partial^*\chi^*_-) \right] -
\nonumber \\
&-& g[ (\partial\phi^{a*})\bar{a}^*_+t^a(\partial^*\chi^*_-)
+ (\partial^*\phi^a)\bar{a}_+t^a(\partial\chi_-) \nonumber \\
&+& (\partial\phi^{a*})(\partial^*\bar{\chi}_-)t^a a_+ +
(\partial^*\phi^a)(\partial\bar{\chi}^*_-)t^aa^*_+ ]
\end{eqnarray}

Let us consider now the contribution (28) for the scattering off
left particles. With the use of mass-shell condition for the
real particles ($\partial_+\partial_- = \partial\partial^*$) we
obtain (cf.~(43))
\begin{eqnarray}
   {\cal L}^{(L)}_{scat} &=& - \frac{ig}{2}A^a_-[
(\partial_+ \partial^*\phi^*) T^a(\partial \phi) + (\partial_+
\partial \phi) T^a(\partial^* \phi^*)] \nonumber \\
  &+& 2gA^a_- [(\partial^*\bar{\chi}_+)t^a(\partial\chi_+)
+ (\partial\bar{\chi}^*_+)t^a(\partial^*\chi^*_+ )] -
\nonumber \\
&-& g[ (\partial\phi^{a})\bar{a}^*_-t^a(\partial^*\chi^*_+)
+ (\partial^*\phi^{a*})\bar{a}_-t^a(\partial\chi_+) \nonumber \\
&+& (\partial\phi^{a})(\partial^*\bar{\chi}_+)t^a a_- +
(\partial^*\phi^{a*})(\partial\bar{\chi}^*_+)t^aa^*_- ]
\end{eqnarray}
\newpage

The production terms of the effective Lagrangian are obtained
from eq.~(26):
\begin{eqnarray}
{\cal L}_{prod} &=& g[ \phi^a (\partial A_-)T^a(\partial^*
A_+) - \phi^{a *}(\partial^* A_-)T^a (\partial A_+) ]
\nonumber \\
&-& \frac{ig}{2}[ \phi^a( - \bar{a}^*_+ t^a(\partial a_-) +
(\partial^*\bar{a}_+)t^a a^*_- ) + \nonumber \\
 &+& \phi^{a *} (\bar{a}_+t^a(\partial^* a^*_-) - (\partial
\bar{a}^*_+)t^a a_-) + \nonumber \\
 &+& \phi^a ( - (\partial \bar{a}^*_-)t^a a_+ + \bar{a}_- t^a
(\partial^* a^*_+)) + \nonumber \\
&+& \phi^{a *}((\partial^*\bar{a}_-)t^a a^*_+ - \bar{a}^*_- t^a
(\partial a_+)) ] + \nonumber \\
&+& ig [ \bar{\chi}_-t^aa^*_-(\partial^*A^a_+) -
\bar{\chi}^*_-t^aa_-(\partial A^a_+) \nonumber \\
&+& \bar{a}_-t^a\chi^*_-(\partial^* A^a_+) - \bar{a}^*_-t^a
\chi_- (\partial A^a_+) \nonumber \\
&+& \bar{a}_+ t^a \chi^*_+ (\partial^* A^a_-) - \bar{a}^*_+ t^a
\chi_+ (\partial A^a_-) \nonumber \\
&+& \bar{\chi}_+ t^a a^*_+ (\partial^* A^a_-) - \bar{\chi}^*_+
t^a a_+(\partial A^a_-) ]
\end{eqnarray}

At last from eqs.~(29),(30) we obtain the following contribution
to the interaction of pure Coulomb fields:
\begin{eqnarray}
{\cal L}_{Coul} &=& \frac{ig}{2} (\partial\partial^*
A^a_-)(\frac{1}{\partial_+}A_+ ) T^a A_+ -
\frac{g}{2}(\frac{1}{\partial_+}A^a_+)[
(\partial \bar{a}^*_-) t^a a_+ + \nonumber \\
&+& (\partial^* \bar{a}_-)t^a a^*_+ + \bar{a}^*_+t^a(\partial
a_-)  + \bar{a}_+t^a (\partial^* a^*_-) ] + \nonumber \\
 &+& \frac{ig}{2} (\partial\partial^*
A^a_+)(\frac{1}{\partial_-}A_- ) T^a A_- -
\frac{g}{2}(\frac{1}{\partial_-}A^a_-)[
(\partial \bar{a}^*_+) t^a a_- + \nonumber \\
&+& (\partial^* \bar{a}_+)t^a a^*_- + \bar{a}^*_-t^a(\partial
a_+)  + \bar{a}_-t^a (\partial^* a^*_+) ]
\end{eqnarray}
Note, that the terms corresponding to the gluon transition into
the quark-anti-quark pair in the $t$-channel are absent in
eq.~(46) because they are small.

As we stressed above the fields $\chi_+$ and $\chi_-$ in the
above action are related by eq.~(41) due to equations of motion.
However one can modify the kinetic action ${\cal
L}^{\chi}_{free}$ for $\chi$ fields ( see eq.~(42)) in the way
corresponding to the initial free fermion contribution $i\bar{\psi}
\gamma^{\mu}\partial_{\mu}\psi$ in eq.~(2)
\begin{eqnarray}
{\cal L}^{\chi}_{free} &=& 4i[
(\partial^*\bar{\chi}_+)(\partial_- \partial\chi_+) +
(\partial\bar{\chi}^*_+)(\partial_-\partial^*\chi^*_+) +
\nonumber \\
&+& (\partial^*\bar{\chi}_-)(\partial_+\partial\chi_-) +
(\partial\bar{\chi}^*_-)(\partial_+\partial^*\chi^*_-) +
\nonumber \\
&+& (\partial^*\bar{\chi}_+)(\partial\partial^*\chi^*_-) +
(\partial\bar{\chi}^*_+)(\partial\partial^*\chi_-) +
\nonumber \\
&+& (\partial^*\bar{\chi}_-)(\partial\partial^*\chi^*_+) +
(\partial\bar{\chi}^*_-)(\partial\partial^*\chi_+) ]
\end{eqnarray}
In this case the constraint (41) will be a consequence of the Dirac
equation for fields $\chi_{\pm}\; ,\; \chi^*_{\pm}$ for the
quark on its mass-shell. The representation (47) for ${\cal
L}^{\chi}_{free}$ is local and has the same symmetry for the
transformation $+ \leftrightarrow -$ as the interaction terms
(43), (44), (45) and (46).
\newpage
\section{Discussion}

We have obtained the effective action for high-energy
multi-Regge processes directly from the original QCD action by
separating the gluon and quark fields into parts with momenta
obeying conditions imposed by the multi-Regge kinematics. The
heavy modes are eliminated by their equations of motion. This
induces new interaction terms. Further it is convenient to
distinguish the Coulomb modes with the kinematics typical for
exchanged particles and the modes corresponding to particles
produced or scattered with small momentum transfer.

We have chosen notations such that the two distinct modes appear
as different fields. However, we have to keep in mind the
momentum ranges appropriate for them. For example, one should
not allow the Coulomb fields ($A_{\pm},\; a_{\pm}$) to have
momenta with $|k_+k_-| \geq |k^2_{\perp}|$, because this would
violate the conditions of multi-Regge kinematics. Also, one
should not allow the scattering fields ($\phi\; ,\;\chi_{\pm}$)
to have momenta with $|k_+k_-| \ll |k^2_{\perp}|$, because this
would double a contribution already accounted for by the Coulomb
fields.

In the present form of the action these restrictions are not
included and are to be taken as external conditions. We are
looking for a modification of the vertices or propagators, which
would suppress the forbidden configurations.

It has been checked that the effective action,
obtained in this paper, reproduces multi-Regge scattering
amplitudes on the tree level. In particular the vertices of
production of gluons and quarks reproduce the known effective
vertices \cite{2}, \cite{12}. The analysis of tree amplitudes
provides a straightforward but fairly heuristic way \cite{9} to
obtain the effective action up to the triple interaction of
Coulomb fields.

The Coulomb fields turn into reggeons by summing over all loops
in LLA. The significance of triple Coulomb interaction depends
on the interpretation of the scattering fields. One can consider
the elimination of the heavy modes as an operation done on the
classical action only ( on mass-shell ). Doing quantization only
after this step would not imply any restriction on the
virtualness of the scattering fields ( besides of avoiding a
mixing with Coulomb fields). Reggeization arises by $s$-channel
interaction and there is no triple Coulomb interaction.  \newline
A different point of view would be to do the elimination of
the heavy modes as an operation done on the functional integral,
i.e. integrating them out approximately. After this the
scattering fields are not allowed to be strongly virtual and
$s$-channel iteration does not give the essential contribution
to reggeization. Here the reggeization arises from the triple
Coulomb interaction.

A triple interaction of negative signature reggeons seems to
contradict the Gribov signature conservation rule. A. White
showed \cite{13} that nevertheless such reggeon interaction can
be treated consistently and that triple-reggeon vertices are
very important for understanding the gluon reggeization.

The derivation outlined here can be made more rigorous by
avoiding the approximation up to the first order in $g$ in the
equations of motion. It provides a deeper understanding of the
effective action. Owing to the approximations and external
conditions the resulting action can be represented in several
forms, some of which emphasize symmetry properties. Note that one
can obtain one-loop corrections to above effective action using
the results of Ref. \cite{14}.

With the effective action we achieved already essential
simplifications compared to the original theory. We consider
this as a first step on the way leading finally to a simple
effective theory from which the leading contributions to the
multi-Reggeon Green functions could be calculated.   \\
$\mbox{ }$ \\
{\Large\bf Acknowledgements} \\
$\mbox{ }$ \\
We would like to thank Prof. H.D. Dahmen and all members of
Theory Department of Siegen University for many discussions and
hospitality.


=====================================


\input feynman


\begin{picture}(24000,20000)
\drawline\fermion[\N\REG](21000,6000)[6000]
\put(21000,4500){$\mbox{p}_{B}$}
\drawarrow[\N\ATTIP](\pmidx,\pmidy)
\drawline\fermion[\N\REG](21000,12000)[6000]
\drawarrow[\N\ATTIP](21000,16500)
\put(21000,19500){$\mbox{p}_{B'}$}
\drawline\fermion[\N\REG](18000,13500)[4500]
\drawarrow[\N\ATTIP](18000,16500)
\put(18000,19500){$\mbox{k}_{n}$}
\drawline\fermion[\N\REG](15000,13500)[4500]
\drawarrow[\N\ATTIP](15000,16500)
\put(15000,19500){$\mbox{k}_{n-1}$}
\drawline\fermion[\N\REG](9000,13500)[4500]
\drawarrow[\N\ATTIP](9000,16500)
\put(9000,19500){$\mbox{k}_{2}$}
\drawline\fermion[\N\REG](6000,13500)[4500]
\drawarrow[\N\ATTIP](6000,16500)
\put(6000,19500){$\mbox{k}_{1}$}
\drawline\fermion[\N\REG](3000,12000)[6000]
\drawarrow[\N\ATTIP](3000,16500)
\put(3000,19500){$\mbox{p}_{A'}$}
\drawline\fermion[\N\REG](3000,6000)[6000]
\drawarrow[\N\ATTIP](3000,9000)
\put(3000,4500){$\mbox{p}_{A}$}
\drawline\fermion[\W\REG](21000,12000)[18000]
\drawarrow[\E\ATTIP](4500,12000)
\drawarrow[\E\ATTIP](7500,12000)
\drawarrow[\E\ATTIP](16500,12000)
\drawarrow[\E\ATTIP](19500,12000)
\put(4300,12700){$\footnotesize q_1$}
\put(7300,12700){$\footnotesize q_2$}
\put(16300,12700){$\footnotesize q_n$}
\put(18600,12700){$\footnotesize q_{n+1}$}
\put(12000,12000){\oval(18000,3000)}
\put(11000,16500){$.\;\; . \;\; .$}
\put(10500,1500){Fig. 1}
\end{picture}

\vspace*{1cm}

\begin{picture}(27000,21000)
\drawline\fermion[\N\REG](3000,3000)[6000]
\drawarrow[\N\ATTIP](3000,6000)
\drawline\fermion[\N\REG](3000,9000)[6000]
\drawarrow[\N\ATTIP](3000,12000)
\drawline\fermion[\N\REG](24000,3000)[6000]
\drawarrow[\N\ATTIP](24000,6000)
\drawline\fermion[\N\REG](24000,9000)[6000]
\drawarrow[\N\ATTIP](24000,12000)
\drawline\fermion[\E\REG](3000,9000)[10500]
\drawarrow[\E\ATTIP](9000,9000)
\drawline\fermion[\E\REG](13500,9000)[10500]
\drawarrow[\E\ATTIP](19500,9000)
\drawline\fermion[\N\REG](13500,9000)[3000]
\drawarrow[\N\ATTIP](13500,10500)
\drawline\fermion[\NE\REG](13500,12000)[4300]
\drawarrow[\NE\ATTIP](\pmidx,\pmidy)
\drawline\fermion[\NW\REG](13500,12000)[4300]
\drawarrow[\NW\ATTIP](\pmidx,\pmidy)
\put(13500,16500){$s_2$}
\put(10500,16000){$\mbox{k}_1$}
\put(16500,16000){$\mbox{k}_2$}
\put(14500,10500){k}
\put(9000,8000){$\mbox{q}_1$}
\put(19500,8000){$\mbox{q}_2$}
\put(12000,500){Fig. 2}
\end{picture}
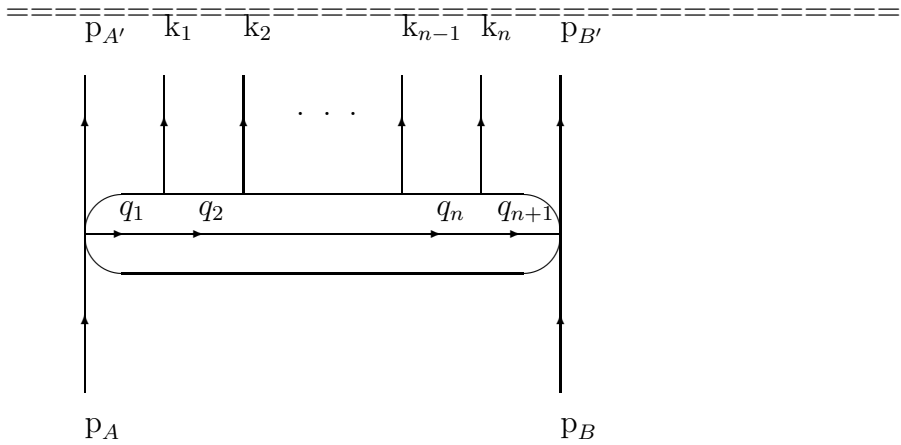
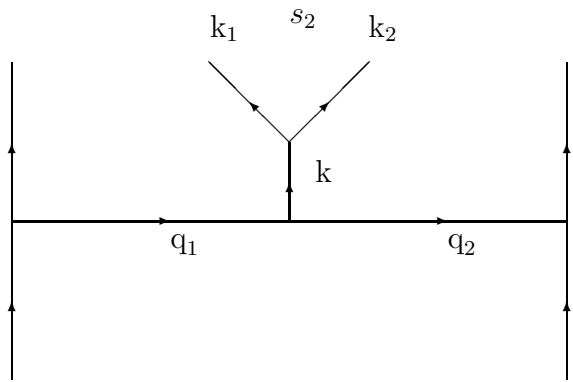
\end{document}